\documentclass[traditabstract]{aa} 
\usepackage{amsmath}
\usepackage{natbib}
\usepackage{graphicx}
\usepackage{color}
\usepackage{txfonts}
\usepackage{subfigure}
\usepackage{xspace}
\usepackage{url}
\newcommand{\reb}{\texttt{REBOUND}\xspace }
\begin{document}
\title{REBOUND: An open-source multi-purpose N-body code for collisional dynamics}

\author{Hanno Rein \inst{1} \and Shang-Fei Liu\inst{2,3} }

\institute{	Institute for Advanced Study,
		1 Einstein Drive, Princeton, NJ 08540\\
		\email{rein@ias.edu}
	\and
		Kavli Institute for Astronomy and Astrophysics, Peking University,
		Beijing 100871, P. R. China
	\and Department of Astronomy, Peking University,
		Beijing 100871, P. R. China\\
	     \email{liushangfei@pku.edu.cn}
	}

\date{Submitted: 13 September 2011 -- Accepted: 6 November 2011}

\abstract{
\reb is a new multi-purpose N-body code which is freely available under an open-source license. 
It was designed for collisional dynamics such as planetary rings but can also solve the classical N-body problem. 
It is highly modular and can be customized easily to work on a wide variety of different problems in astrophysics and beyond.

\reb comes with three symplectic integrators: leap-frog, the symplectic epicycle integrator (SEI) and a Wisdom-Holman mapping (WH). 
It supports open, periodic and shearing-sheet boundary conditions.
\reb can use a Barnes-Hut tree to calculate both self-gravity and collisions. 
These modules are fully parallelized with MPI as well as OpenMP. 
The former makes use of a static domain decomposition and a distributed essential tree.
Two new collision detection modules based on a plane-sweep algorithm are also implemented.
The performance of the plane-sweep algorithm is superior to a tree code for simulations in which one dimension is much longer than the other two and in simulations which are quasi-two dimensional with less than one million particles.

In this work, we discuss the different algorithms implemented in \reb, the philosophy behind the code's structure as well as implementation specific details of the different modules. 
We present results of accuracy and scaling tests which show that the code can run efficiently on both desktop machines and large computing clusters.
}

\keywords{ Methods: numerical -- Planets and satellites: rings --  Proto-planetary disks }

\maketitle

\section{Introduction}
\reb is a new open-source collisional N-body code.
This code, and precursors of it, have already been used in wide variety of publications \citep[][Rein \& Liu in preparation; Rein \& Latter in preparation]{ReinPapaloizou2010,Crida2010,ReinLesurLeinhardt2010}. 
We believe that \reb can be of great use for many different problems and have a wide reach in astrophysics and other disciplines. 
To our knowledge, there is currently no publicly available code for collisional dynamics capable of solving the problems described in this paper.
This is why we decided to make it freely available under the open-source license GPLv3\footnote{The full license is distributed together with \reb. It can also be downloaded from \url{http://www.gnu.org/licenses/gpl.html}.}. 

Collisional N-body simulations are extensively used in astrophysics.
A classical application is a planetary ring \citep[see e.g.][and references therein]{Wisdom1988,Salo1991,Richardson1994,LewisStewart2009,ReinPapaloizou2010,MichikoshiKokubo2011} which have often a collision time-scale that is much shorter than or at least comparable to an orbital time-scale.
Self-gravity plays an important role, especially in the dense parts of Saturn's rings \citep{Schmidt2009}. 
These simulations are usually done in the shearing sheet approximation \citep{Hill1878}. 

Collisions are also important during planetesimal formation \citep[][Johansen et al. in preparation]{Johansen2007,ReinLesurLeinhardt2010}. 
Collisions provide the dissipative mechanism to form a planetesimal out of a gravitationally bound swarm of boulders. 

\reb can also be used with little modification in situations where only a statistical measure of the collision frequency is required such as in transitional and debris discs.
In such systems, individual collisions between particles are not modeled exactly, but approximated by the use of super-particles \citep{StarkKuchner2009,LithwickChiang2007}. 

Furthermore, \reb can be used to simulate classical N-body problems involving entirely collision-less systems. 
A symplectic and mixed variable integrator can be used to follow the trajectories of both test-particles and massive particles.

We describe the general structure of the code, how to obtain, compile and run it in Sect.~\ref{sec:overview}.
The time-stepping scheme and our implementation of symplectic integrators are described in Sect.~\ref{sec:integrators}.
The modules for gravity are described in Sect.~\ref{sec:gravity}.
The algorithms for collision detection are discussed in Sect.~\ref{sec:collisions}.
In Sect.~\ref{sec:tests}, we present results of accuracy tests for different modules.
We discuss the efficiency of the parallelization with the help of scaling tests in Sect.~\ref{sec:scaling}.
We finally summarize in Sect.~\ref{sec:summary}.

\section{Overview of the code structure}\label{sec:overview}
\reb is written entirely in C and conforms to the ISO C99 standard. 
It compiles and runs on any modern computer platform which supports the POSIX standard such as Linux, Unix and Mac OSX. 
In its simplest form, \reb requires no external libraries to compile. 

Users are encouraged to install the OpenGL and GLUT libraries which enable real-time and interactive 3D visualizations. 
LIBPNG is required to automatically save screen-shots. 
The code uses OpenMP for parallelization on shared memory systems. 
Support for OpenMP is built-in to modern compilers and requires no libraries (for example gcc~$\ge 4.2$).
An MPI library must be installed for parallelization on distributed memory systems.
\reb also supports hybrid parallelization using both OpenMP and MPI simultaneously.

\subsection{Downloading and compiling the code}
The source code is hosted on the github platform and can be downloaded at \url{http://github.com/hannorein/rebound/}. 
A snapshot of the current repository is provided as tar and zip-files. 
However, users are encouraged to clone the entire repository with the revision control system \texttt{git}. 
The latter allows one to keep up-to-date with future updates.
Contributions from users to the public repository are welcome.
Once downloaded and extracted, one finds five main directories. 

The entire source code can be found in the \texttt{src/} directory. 
In the vast majority of cases, nothing in this directory needs to be modified. 

Many examples are provided in the \texttt{examples/} directory. 
Each example comes with a problem file, named \texttt{problem.c}, and a makefile named \texttt{Makefile}. 
To compile one of the examples, one has to run the \texttt{make} command in the corresponding directory. 
The code compilation is then performed in the following steps:
\begin{enumerate}
\item The makefile sets up environment variables which control various options such as the choice of compiler, code optimization, real time visualization and parallelization.
\item It sets symbolic links, specifying the modules chosen for this problem (see below). 
\item It calls the makefile in the \texttt{src/} directory which compiles and links all source files. 
\item The binary file is copied to the problem directory, from where it can be executed.
\end{enumerate}

Documentation of the source code can be generated in the \texttt{doc/} directory with doxygen.
There is no static documentation available because the code structure depends crucially on the modules currently selected.
To update the documentation with the current module selection, one can simply run \texttt{make doc} in any directory with a makefile.

In the directory \texttt{tests/} one finds tests for accuracy and scaling as well as simple unit tests. 
The source code of the tests presented in Sects.~\ref{sec:tests} and~\ref{sec:scaling} is included as well.

The \texttt{problem/} directory is the place to create new problems. It contains a template for that. 
Any of the examples can also be used as a starting point for new problems.

\subsection{Modules}
\reb is extremely modular. 
The user has the choice between different gravity, collision, boundary and integration modules. 
It is also possible to implement completely new modules with minimal effort. 

Modules are chosen by setting symbolic links. 
Thus, there is no need to execute a configuration script prior to compiling the code.
For example, there is one link \texttt{gravity.c} which points to one of the gravity modules \texttt{gravity\_*.c}. 
The symbolic links are set in each problem makefile.
Only this makefile has to be changed when a different module is used.
Pre-compiler macros are set automatically for situations in which different modules need to know about each other.

This setup allows the user to work on multiple projects at the same time using different modules.
When switching to another problem, nothing has to be set-up and the problem can by compiled by simply typing \texttt{make} in the corresponding directory.

To implement a new module, one can just copy an existing module to the problem directory, modify it and change the link in the makefile accordingly. 
Because no file in the \texttt{src/} directory needs to be changed, one can easily keep \reb in sync with new versions\footnote{On how to do that, see for example \url{http://gitref.org/} for an introduction to git.}.

\subsection{Computational domain and boundary conditions}\label{sec:domain}
In \reb, the computational domain consists of a collection of cubic boxes. 
Any integer number of boxes can be used in each direction. 
This allows elongated boxes to be constructed out of cubic boxes.
The cubic root boxes are also used for static domain decomposition when MPI is enabled.
In that case, the number of root boxes has to be a integer multiple of the number of MPI nodes. 
When a tree is used for either gravity or collision detection, there is one tree structure per root box (see Sect.~\ref{sec:gravitytree}). 

\reb comes with three different boundary conditions. 
Open boundaries (\texttt{boundaries\_open.c}) remove every particle from the simulation that leaves the computational domain. 
Periodic boundary conditions (\texttt{boundaries\_periodic.c}) are implemented with ghost boxes.
Any number of ghost boxes can be used in each direction. 
Shear-periodic boundary conditions (\texttt{boundaries\_shear.c}) can be used to simulate a shearing sheet.

\section{Integrators}\label{sec:integrators}
Several different integrators have been implemented in \reb.
Although all of these integrators are second order accurate and symplectic, their symplectic nature is formally lost as soon as self-gravity or collisions are approximated or when velocity dependent forces are included. 

All integrators follow the commonly used Drift-Kick-Drift (DKD) scheme\footnote{We could have also chosen a KDK scheme but found that a DKD scheme performs slightly better.} but implement the three sub-steps differently.
We describe the particles' evolution in terms of a Hamiltonian $H$ which can often be written as the sum of two Hamiltonians $H=H_1+H_2$. 
How the Hamiltonian is split into two parts depends on the integrator.
Usually, one identifies $H_1(p)$ as the kinetic part and $H_2(q)$ as the potential part, where $p$ and $q$ are the canonical momenta and coordinates.
During the first drift sub-step, the particles evolve under the Hamiltonian $H_1$ for half a time-step $dt/2$.
Then, during the kick sub-step, the particles evolve under the Hamiltonian $H_2$ for a full time-step $dt$. 
Finally, the particles evolve again for half a time-step under $H_1$.
Note that the positions and velocities are synchronized in time only at the end of the DKD time-steps.
We refer the reader to \cite{SahaTremaine1992} and references therein for a detailed discussion on symplectic integrators.

\reb uses the same time-step for all particles. 
By default, the time-step does not change during the simulation because in all the examples that come with \reb, the time-step can be naturally defined as a small fraction of the dynamical time of the system.
However, it is straight forward to implement a variable time-step.
This implementation depends strongly on the problem studied.
Note that in general variable time-steps also break the symplectic nature of an integrator.

\reb does not choose the time-step automatically. 
It is up to the user to ensure that the time-step is small enough to not affect the results.
This is especially important for highly collisional systems in which multiple collisions per time-step might occur and in situations where the curvature of particle trajectories is large.
The easiest way to ensure numerical convergence is to run the same simulation with different time-steps.
We encourage users to do that whenever a new parameter regime is studied.

\subsection{Leap-frog}
Leap-frog is a second-order accurate and symplectic integrator for non-rotating frames. 
Here, the Hamiltonian is split into the kinetic part $H_1=\frac12p^2$ and the potential part $H_2=\Phi(x)$.
Both the drift and kick sub-steps are simple Euler steps.
First the positions of all particles are advanced for half a time-step while keeping the velocities fixed. 
Then the velocities are advanced for one time-step while keeping the positions fixed.
In the last sub-step the velocities are again advanced for half a time-step.
Leap-frog is implemented in the module \texttt{integrator\_leapfrog.c}.

\subsection{Wisdom-Holman Mapping}
A symplectic Wisdom-Holman mapping \citep[WH,][]{WisdomHolman1991} is implemented as a module in \texttt{integrator\_wh.c}. 
The implementation follows closely that by the SWIFT code\footnote{\url{http://www.boulder.swri.edu/~hal/swift.html}}.
The WH mapping is a mixed variable integrator that calculates the Keplerian motion of two bodies orbiting each other exactly up to machine precision during the drift sub-step. 
Thus, it is very accurate for problems in which the particle motion is dominated by a central $1/r$ potential and perturbations added in the kick sub-step are small.
However, the WH integrator is substantially slower than the leap-frog integrator because Kepler's equation is solved iteratively every time-step for every particle.

The integrator assumes that the central object has the index~$0$ in the particle array, that it is located at the origin and that it does not move. 
The coordinates of all particles are assumed to be the heliocentric frame. 
During the sub-time-steps the coordinates are converted to Jacobi coordinates (and back) according to their index.
The particle with index~$1$ has the first Jacobi index, and so on.
This works best if the particles are sorted according to their semi-major axis.
Note that this is not done automatically. 

\subsection{Symplectic Epicycle Integrator}
The symplectic epicycle integrator \citep[SEI,][]{ReinTremaine2011} for Hill's approximation \citep{Hill1878} is implemented in \texttt{integrator\_sei.c}. When shear-periodic boundary conditions (\texttt{boundaries\_shear.c}) are used, the Hill approximation is know as a shearing sheet. 

SEI has similar properties to the Wisdom-Holman mapping in the case of the Kepler potential but works in a rotating frame and is as fast as a standard non-symplectic integrator. 
The error after one time-step scales as the third power of the time-step times the ratio of the gravitational force over the Coriolis force \citep[see][for more details on the performance of SEI]{ReinTremaine2011}.

The epicyclic frequency $\Omega$ and the vertical epicyclic frequency $\Omega_z$ can be specified individually. 
This can be used to enhance the particle density in the mid-plane of a planetary ring and thus simulate the effect of self-gravity \citep[see e.g.][]{Schmidtetal2001}.

\section{Gravity}\label{sec:gravity}
\reb is currently equipped with two (self)-gravity modules.  
A gravity module calculates exactly or approximately the acceleration onto each particle.
For a particle with index $i$ this is given by
\begin{align}
\mathbf{a}_i &= \sum_{j=0}^{N_\mathrm{active}-1} \frac{Gm_j}{\left(r_{ji}^2+b^2\right)^{3/2}}\; \mathbf{\hat r_{ji}},  \label{eq:selfgravity}
\end{align}
where $G$ is the gravitational constant, $m_j$ the mass of particle~$j$ and $\mathbf{r}_{ji}$ the relative distance between particles~$j$ and~$i$. 
The gravitational softening parameter $b$ defaults to zero but can be set to a finite value in simulations where physical collisions between particles are not resolved and close encounters might lead to large unphysical accelerations.
The variable $N_\mathrm{active}$ specifies the number of massive particles in the simulation. 
Particles with an index equal or larger than $N_\mathrm{active}$ are treated as test-particles.
By default, all particles are assumed to have mass and contribute to the sum in Eq.~\ref{eq:selfgravity}.

\subsection{Direct summation}
The direct summation module is implemented in \texttt{gravity\_direct.c} and computes Eq.~\ref{eq:selfgravity} directly. 
If there are $N_\mathrm{active}$ massive particles and $N$ particles in total, the performance scales as $\mathcal{O}(N\cdot N_\mathrm{active})$.
Direct summation is only efficient with few active particles; typically $N_\mathrm{active} \lesssim10^2$.

\subsection{Octree} \label{sec:gravitytree}
\citet[][BH hereafter]{Barnes1986} proposed an algorithm to approximate Eq.~\ref{eq:selfgravity}, which can reduce the computation time drastically from $\mathcal{O}(N^2)$ to $\mathcal{O}(N \log N)$. 
The idea is straightforward: distant particles contribute to the gravitational force less than those nearby. 
By grouping particles hierarchically, one can separate particles in being far or near from any one particle.
The total mass and the center of mass of a group of particles which are far away can then be used as an approximation when calculating the long-range gravitational force.
Contributions from individual particles are only considered when they are nearby. 

We implement the BH algorithm in the module \texttt{gravity\_tree.c}.
The hierarchical structure is realized using a three-dimensional tree, called an octree.
Each node represents a cubic cell which might have up to eight sub-cells with half the size. 
The root node of the tree contains all the particles, while the leaf nodes contain exactly one particle. 
The BH tree is initially constructed by adding particles one at a time to the root box, going down the tree recursively to smaller boxes until one reaches an empty leaf node to which the particle is then added.
If the leaf node already contains a particle it is divided into eight sub-cells. 

Every time the particles move, the tree needs to be updated using a tree reconstruction algorithm.
We therefore keep track of any particle crossing the boundaries of the cell it is currently in.
If it has moved outside, then the corresponding leaf node is destroyed and the particle is re-added to the tree as described above. 
After initialization and reconstruction, we walk through the tree to update the total mass and the center of mass for each cell from the bottom-up.

To calculate the gravitational forces on a given particle, one starts at the root node and descends into sub-cells as long as the cells are considered to be close to the particle. 
Let us define the opening angle as $\theta = w/R$, where $w$ is the width of the cell and $R$ is the distance from the cell's center of mass to the particle.
If the opening angle is smaller than a critical angle $\theta_{\mathrm{crit}}>\theta$, the total mass and center of mass of the cell are used to calculate the contribution to the gravitational force. 
Otherwise, the sub-cells are opened until the criterion is met. 
One has to choose $\theta_{\mathrm{crit}}$ appropriately to achieve a balance between accuracy and speed. 

\reb can also include the quadrupole tensor of each cell in the gravity calculation by setting the pre-compiler flag \texttt{QUADRUPOLE}. 
The quadrupole expansion \citep{Hernquist1987} is more accurate but also more time consuming for a fixed $\theta_{\mathrm{crit}}$.
We discuss how the critical opening angle and the quadrupole expansion affect the accuracy in Sect.~\ref{sec:forceaccuracy}.

\begin{figure}
\centering \resizebox{0.9\columnwidth}{!}{\input{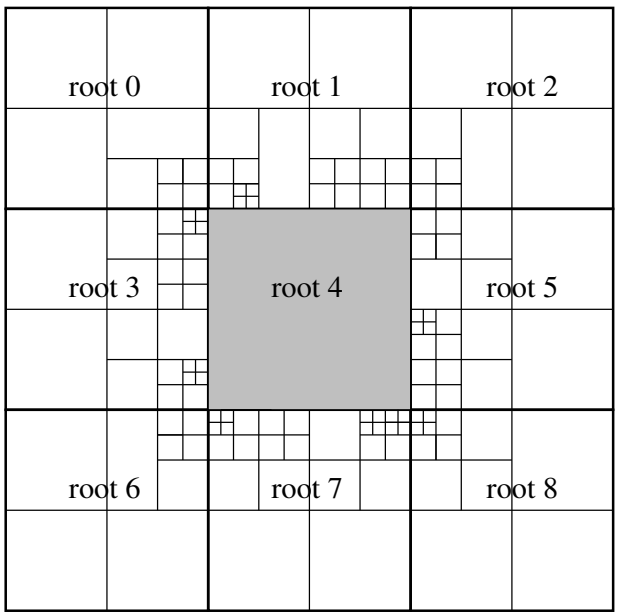}}
\caption{Illustration of the essential trees needed by root box 4. 
The different levels of the tree structure which need to be copied depend on the distance to the nearest boundary of root box 4 and the opening angle $\theta$. 
See text for details.
\label{fig:essentialtree}}
\end{figure}

With \reb, a static domain decomposition is used for parallelizing the tree algorithm on distributed memory systems.
Each MPI node contains one or more root boxes (see also Sect.~\ref{sec:domain}) and all particles within these boxes belong to that node.
The number of root boxes $N_{\mathrm{RB}}$ has to be a multiple of the number of MPI nodes $N_{\mathrm{MPI}}$.
For example, the setup illustrated in Fig.~\ref{fig:essentialtree} uses 9 root boxes allowing 1, 3 or 9~MPI~nodes.
By default, the domain decomposition is done first along the $z$ direction, then along the $y$ and $x$ direction. 
If one uses 3 MPI~nodes in the above example, the boxes $0-2$ are on on node~$0$, the boxes $3-5$ on node~$1$ and the remaining boxes on node~$2$.
When a particle moves across a root box boundary during the simulation, it is send to the corresponding node and removed form the local tree.

Because of the long-range nature of gravity, every node needs information from any other node during the force calculation. 
We distribute this information before the force calculation using an essential tree \citep{Salmon1990} and an all-to-all communication pattern.
The essential tree contains only those cells of the local tree that might be accessed by the remote node during the force calculation.
Each node prepares a total of $N_\mathrm{RB}-N_\mathrm{RB}/N_\mathrm{MPI}$ different essential trees.
The cells that constitute the essential tree are copied into a buffer array and the daughter cell references therein are updated accordingly.
The center of mass and quadrupole tensors (if enabled) are stored in the cell structure and automatically copied when a cell is copied.
For that reason only the tree structure needs to be distributed, not individual particles.
The buffer array is then sent to the other nodes using non-blocking MPI calls.

For example, suppose 9 MPI~nodes are used, each node using exactly one tree in its domain.
For that scenario the essential trees prepared for root box~$4$ are illustrated in Fig.~\ref{fig:essentialtree}.
The essential trees include all cells which are close enough to the boundary of root box~$4$ so that they might be opened during the force calculation of a particle in root box~$4$ according to the opening angle criteria.

In Sect.~\ref{sec:scaling} we show that this parallelization is very efficient when the particle distribution is homogeneous and there are more than a few thousand particles on every node. 
When the number of particles per node is small, communication between nodes dominates the total runtime.

\section{Collisions}\label{sec:collisions}
\begin{figure}
\centering \resizebox{0.9\columnwidth}{!}{\input{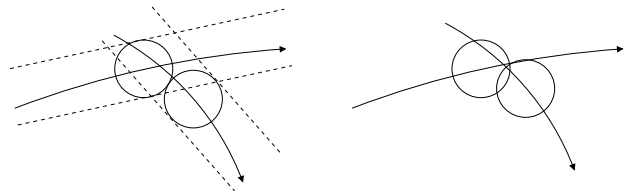}}
\caption{Different collision detection algorithms. Left: curved particle trajectories are approximated by straight lines. Right: trajectories are not approximated, particles only collide when they are overlapping. See text for details.\label{fig:collisions}}
\end{figure}
\reb supports several different modules for collision detection which are described in detail below. 
All of these methods search for collisions only approximately, might miss some of the collisions or detect a collision where there is no collision. 
This is because either curved particle trajectories are approximated by straight lines (\texttt{collisions\_sweep.c} and \texttt{collisions\_sweepphi.c}) or particles have to be overlapping to collide (\texttt{collisions\_direct.c} and \texttt{collisions\_tree.c}). 
This is also illustrated in Fig.~\ref{fig:collisions}. 

In all modules, the order of the collisions is randomized.
This ensures that there is no preferred ordering which might lead to spurious correlations when one particles collides with multiple particles during one time-step.
Note that \reb uses a fixed time-step for all particles. 
Therefore one has to ensure that the time-step is chosen small enough so that one particle does collide with no more than one other particle during one time-step, at least on average. 
See also the discussion in Sect.~\ref{sec:integrators}.

A free-slip, hard-sphere collision model is used. Individual collisions are resolved using momentum and energy conservation.
A constant or an arbitrary velocity dependent normal coefficient of restitution $\epsilon$ can be specified to model inelastic collisions.
The relative velocity after one collision is then given by
\begin{align}
\begin{array}{l}
v_n' = -\epsilon \,v_n\\
v_t' = v_t,
\end{array}
\end{align}
where $v_n$ and $v_t$ are the relative normal and tangential velocities before the collision.
Particle spin is currently not supported.

\subsection{Direct nearest neighbor search}
A direct nearest neighbor collisions search is the simplest collision module in \reb. 
It is implemented in \texttt{collisions\_direct.c},

In this module, a collision is detected whenever two particles are overlapping at the end of the DKD~time-step, i.e. the middle of the drift sub-step, where positions and velocities are synchronized in time (see Sect.~\ref{sec:integrators}). 
This is illustrated in the right panel of Fig.~\ref{fig:collisions}. 
Then, the collision is added to a collision queue. 
When all collisions have been detected, the collision queue is shuffled randomly. 
Each individual collision is then resolved after checking that the particles are approaching each other. 

Every pair of particles is checked once per time-step, making the method scale as $\mathcal{O}(N^2)$. 
Similar to the direct summation method for gravity, this is only useful for a small number of particles. 
For most cases, the nearest neighbor search using  a tree is much faster (see next section).

\subsection{Octree}
The octree described in Sect.~\ref{sec:gravitytree} can also be used to search for nearest neighbors.
The module \texttt{collisions\_tree.c} implements such a nearest neighbor search.
It is parallelized with both OpenMP and MPI.
It can be used in conjunction with any gravity module, but when both tree modules \texttt{gravity\_tree.c} and \texttt{collisions\_tree.c} are used simultaneously, only one tree structure is needed.
When \texttt{collisions\_tree.c} is the only tree module, center of mass and octopole tensors are not calculated in tree cells. 

To find overlapping particles for particle $i$, one starts at the root of the tree and descents into daughter cells as long as the distance of the particle to the cell center $r_{ic}$ is smaller than a critical value:
\begin{align}
r_{ic} &<  R_i + R_{max} + \frac{\sqrt 3}2 w_c,\label{eq:essentialtreecollisions}
\end{align}
where $R_i$ is the size of the particle, $R_{max}$ is the maximum size of a particle in the simulation and $w_c$ is the width of the current cell. 
When two particles are found to be overlapping, a collision is added to the collision queue and resolved later in the same way as above.

If MPI is used, each node prepares the tree and particle structures that are close to the domain boundaries as these might be needed by other nodes (see Fig. \ref{fig:essentialtree}). 
This essential tree is send to other nodes and temporarily added to the local tree structure.
The nearest neighbor search can then be performed in the same way as in the serial version. 
The essential tree and particles are never modified on a remote node.

This essential tree is different from the essential tree used for the gravity calculation in two ways.
First, this tree is needed at the end of the time-step, whereas the gravity tree is needed at the beginning of the kick sub time-step.
Second, the criteria for cell opening, Eq.~\ref{eq:essentialtreecollisions}, is different.

A nearest neighbor search using the octree takes on average $\mathcal{O}(\log(N))$ operations for one particle and therefore $\mathcal{O}(N\log(N))$ operations for all $N$ particles.

\subsection{Plane-sweep Algorithm}
We further implement two collision detection modules based on a plane-sweep algorithm in \texttt{collisions\_sweep.c} and \texttt{collisions\_sweepphi.c}. 
The plane-sweep algorithm makes use of a conceptual plane that is moved along one dimension. 

The original algorithm described by \cite{BentleyOttmann1979} maintains a binary search tree in the orthogonal dimensions and keeps track of line crossings.
In our implementation, we assume the dimension normal to the plane is much longer than the other dimensions.
This allows us to simplify the Bentley-Ottmann algorithm and get rid of the binary search tree which further speeds up the calculation.

In \reb the sweep is either performed along the $x$-direction or along the azimuthal angle $\phi$ (measured in the $xy$-plane from the origin). 
The sweep in the $x$ direction can also be used in the shearing sheet.
The sweep in the $\phi$ direction is useful for (narrow) rings in global simulations.
Here, we only discuss the plane-sweep algorithm in the Cartesian case (along the $x$-direction) in detail. 
The $\phi$~sweep implementation is almost identical except of the difference in periodicity and the need to calculate the angle and angular frequency for every particle at the beginning of the collision search.

Our plane-sweep algorithm can be described as follows (see also Fig.~\ref{fig:planesweep}):
\begin{figure}
\centering \resizebox{0.9\columnwidth}{!}{\input{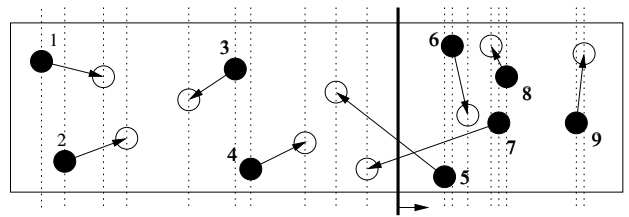}}
\caption{Illustration of the plane-sweep algorithm. The plane is intersecting the trajectories of particles~5 and~7. See text for details.\label{fig:planesweep}}
\end{figure}

\begin{figure*}
\centering 
\subfigure[Force accuracy as a function of the opening angle $\theta_{\mathrm{crit}}$.]{
\centering \resizebox{0.99\columnwidth}{!}{\includegraphics{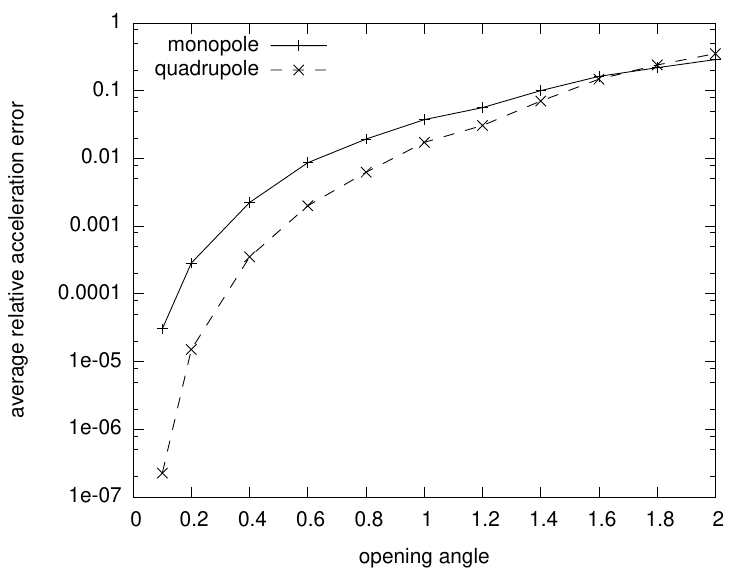}\label{fig:accuracyforce}}
}
\subfigure[Force accuracy as a function of the computation time.]{
\centering \resizebox{0.99\columnwidth}{!}{\includegraphics{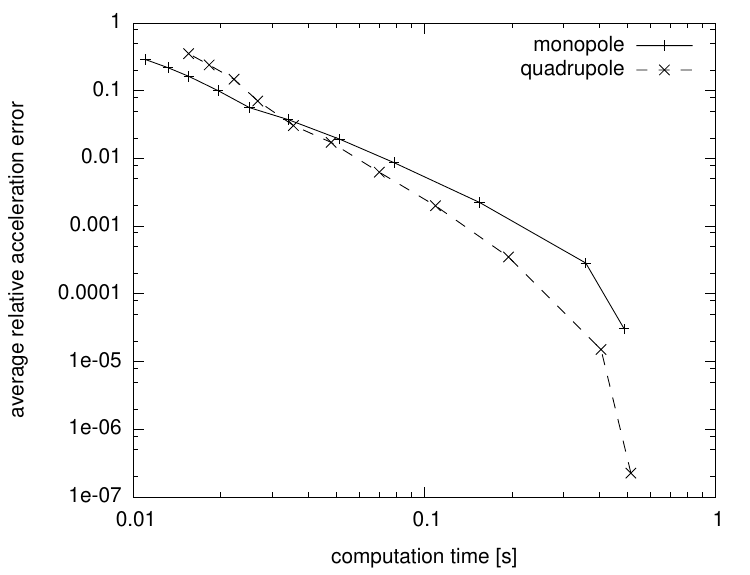}\label{fig:accuracyforce2}}
}
\caption{Comparison of the monopole and quadrupole expansion.}
\end{figure*}

\begin{enumerate}
\item If a tree is not used to calculate self-gravity, the particles are sorted according to their $x$ coordinate\footnote{Each tree cell keeps a reference to the particle it contains. This reference has to be updated every time a particle is moved in the particle array which would lead to larger overhead.}. 
During the first time-step, quicksort is used as the particles are most likely not pre-sorted. 
In subsequent time-steps, the adaptive sort algorithm insertionsort is used.
It can make use of the pre-sorted array from the previous time-step and has an average performance of $\mathcal{O}(N)$ as long as particles do not completely randomize their positions in one time-step.

\item The $x$ coordinate of every particle before and after the drift step is inserted into an array \texttt{SWEEPX}. 
The trajectory is approximated by a line (see left panel of Fig.~\ref{fig:collisions}). 
In general, the real particle trajectories will be curved. 
In that case the positions are only approximately the start and end points of the particle trajectory.
The particle radius is subtracted/added to the minimum/maximum $x$ coordinate. 
The array contains $2N$ elements when all particles have been added. 

\item If a tree is not used, the array \texttt{SWEEPX} is sorted with the $x$ position as a key using the insertionsort algorithm. 
Because the particle array is pre-sorted, insertionsort runs in approximately $\mathcal{O}(N)$ operations. 
If a tree is used, the array is sorted with quicksort.

\item A conceptual plane with its normal vector in the $x$ direction is inserted at the left side of the box. 
While going through the array \texttt{SWEEPX}, we move the plane towards the right one step at a time according to the $x$ coordinate of the current element in the array. 
We thus move the plane to the other side of the box in a total of $2N$ stops.

\item The plane is intersecting particle trajectories. 
We keep track of these intersection using a separate array \texttt{SWEEPL}. 
Whenever a particle appears for the first time in the array \texttt{SWEEPX} the particle is added to the \texttt{SWEEPL} array. 
The particle is removed from the array \texttt{SWEEPL} when it appears in the array \texttt{SWEEPX} for the second time. 
In Fig.~\ref{fig:planesweep}, the plane is between stop 10 and 11, intersecting the trajectories of particles~5 and~7. 

\item When a new particle is inserted into the array \texttt{SWEEPL}, we check for collisions of this particle with any other particle in \texttt{SWEEPL} during the current time-step. 
The collision is recorded and resolved later.
In Fig.~\ref{fig:planesweep} the array \texttt{SWEEPL} has two entries, particles~5 and~7. 
Those will be checked for collisions. 

\end{enumerate}
The time needed to search for a collision at each stop of the plane is $\mathcal{O}(N_{\mathtt{SWEEPL}})$, where $N_{\mathtt{SWEEPL}}$ is the number of elements in the array \texttt{SWEEPL}. 
This could be reduced with a binary search tree to $\mathcal{O}(\log(N_{\mathtt{SWEEPL}}))$ as in the original algorithm by \cite{BentleyOttmann1979}.
However tests have shown that there is little to no performance gain for the problems studied with \reb because a more complicated data structure has to be maintained. 
One entire time-step with the plane-sweep algorithm is completed in $\mathcal{O}(N\cdot N_{\mathtt{SWEEPL}})$.
It is then easy to see that this method can only be efficient when $N_{\mathtt{SWEEPL}} \lesssim \log(N)$, as a tree code is more efficient otherwise.

Indeed, experiments have shown (see Sect.~\ref{sec:scalingcollisions}) that the plane-sweep algorithm is more efficient than a nearest neighbor search with an octree by many orders of magnitude for low dimensional systems in which $N_{\mathtt{SWEEPL}}$ is small.

\section{Test problems}\label{sec:tests}
We present several tests in this section which verify the implementation of all modules.
First, we measure the accuracy of the tree code.
Then we check for energy and momentum conservation.
We use a long term integration of the outer solar system as a test of the symplectic WH integrator. 
Finally, we measure the viscosity in a planetary ring which is a comprehensive test of both self-gravity and collisions. 

\subsection{Force accuracy}\label{sec:forceaccuracy}

We measure the accuracy of the tree module \texttt{gravity\_tree.c} by comparing the force onto each particle to the exact result obtained by direct summation (Eq.~\ref{eq:selfgravity}). 
We set up 1000 randomly distributed particles with different masses in a box.
We do not use any ghost boxes and particles do not evolve. 

We sum up the absolute value of the acceleration error for each particle and normalize it with respect to the total acceleration \citep[see][for more details]{Hernquist1987}. 

This quantity is plotted as a function of the critical opening angle $\theta_{\mathrm{crit}}$ in Fig.~\ref{fig:accuracyforce}.
One can see that the force quickly converges towards the correct value for small $\theta_{\mathrm{crit}}$. 
The quadrupole expansion performs one order of magnitude better then the monopole expansion for $\theta_{\mathrm{crit}}\sim0.5$ and two orders of magnitude better for $\theta_{\mathrm{crit}}\sim0.1$. 

In Fig.~\ref{fig:accuracyforce2} we plot the errors of the same simulations as a function of the computation time.
The quadrupole expansion requires more CPU time than the monopole expansion for fixed $\theta_{\mathrm{crit}}$.
However, the quadrupole expansion is faster when $\theta_{\mathrm{crit}}\lesssim 1$ for a fixed  accuracy.
Note that including the quadrupole tensor also increases communication costs between MPI nodes.

\subsection{Energy and momentum conservation in collisions}
In a non-rotating simulation box with periodic boundaries and non-gravitating collisional particles, we test both momentum and energy conservation.
Using a coefficient of restitution of unity (perfectly elastic collisions), the total momentum and energy is conserved up to machine precision for all collision detection algorithms.

\subsection{Long term integration of Solar System}
To test the long-term behavior of our implementation of the Wisdom-Holman Mapping, we integrate the outer Solar System for 200~million years. 
We use the initial conditions given by \cite{Applegate1986} with 4 massive planets and Pluto as a test particle.
The direct summation module has been used to calculate self-gravity.
With a 40~day time-step and an integration time of $200$~Myr, the total runtime on one CPU was less then 2 hours. 

\begin{figure}
\centering \resizebox{0.99\columnwidth}{!}{\includegraphics{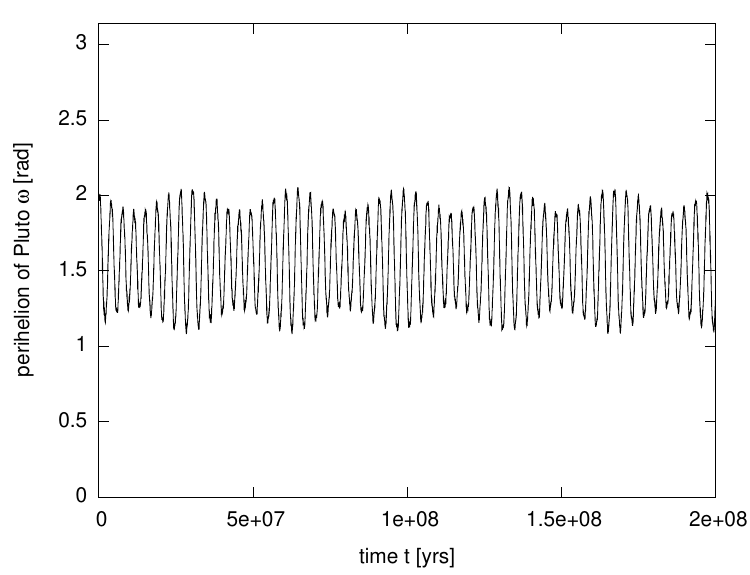}}
\caption{Libration pattern of Pluto with two distinct frequencies of $3.8$~Myr and $34$~Myr.\label{fig:pluto}}
\end{figure}
In Fig.~\ref{fig:pluto}, we plot the perihelion of Pluto as a function of time. 
One can clearly see two distinct libration frequencies with $3.8$~Myr and $34$~Myr time-scales respectively. 
This is in perfect agreement with \cite{Applegate1986}. 

\subsection{Viscosity in planetary rings}
\cite{Daisaka2001} calculate the viscosity in a planetary ring using numerical simulations.
We repeat their analysis as this is an excellent code test as the results depend on both self-gravity and collisions.
The quasi-equilibrium state is dominated by either self-gravity or collisions, depending on the ratio of the Hill radius over the physical particle radius, $r_h^\star$.

\begin{figure}
\centering \resizebox{0.99\columnwidth}{!}{\includegraphics{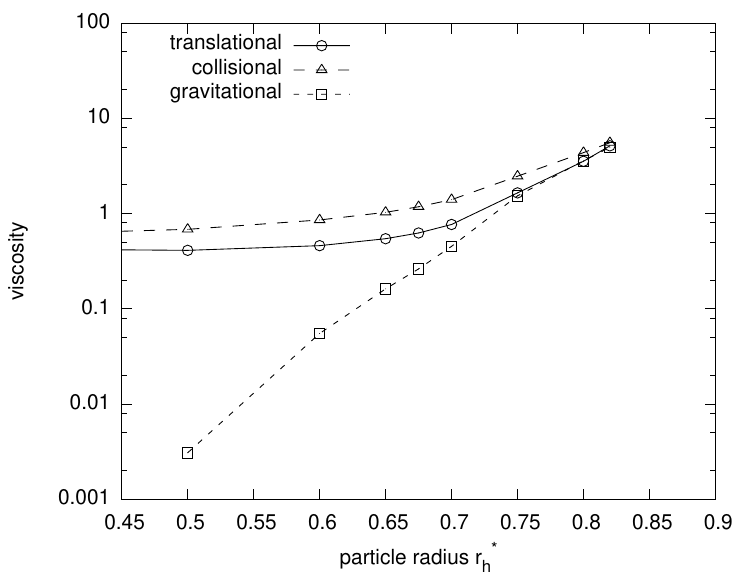}}
\caption{Individual components of the viscosity as a function of the non-dimensional particle radius.\label{fig:viscosity}}
\end{figure}
In this simulation we use the octree implementation for gravity and the plane-sweep algorithm for collisions. 
The geometric optical depth is $\tau=0.5$ and we use a constant coefficient of restitution of $\epsilon=0.5$.
The separate parts of the viscosity are calculated directly as defined by \cite{Daisaka2001} for various $r_h^\star$ and plotted in dimensionless units in Fig.~\ref{fig:viscosity}.

The results are in good agreement with previous results. 
At large $r_h^\star$, the collisional part of the viscosity is slightly higher in our simulations when permanent particle clumps form.
This is most likely due to the the different treatment of collisions and some ambiguity in defining the collisional viscosity when particles are constantly touching each other (Daisaka, private communication).

\section{Scaling}\label{sec:scaling}
\begin{figure*}
\centering 
\subfigure[Strong scaling test: constant problem size, varying number of nodes.]{
\resizebox{0.99\columnwidth}{!}{\includegraphics{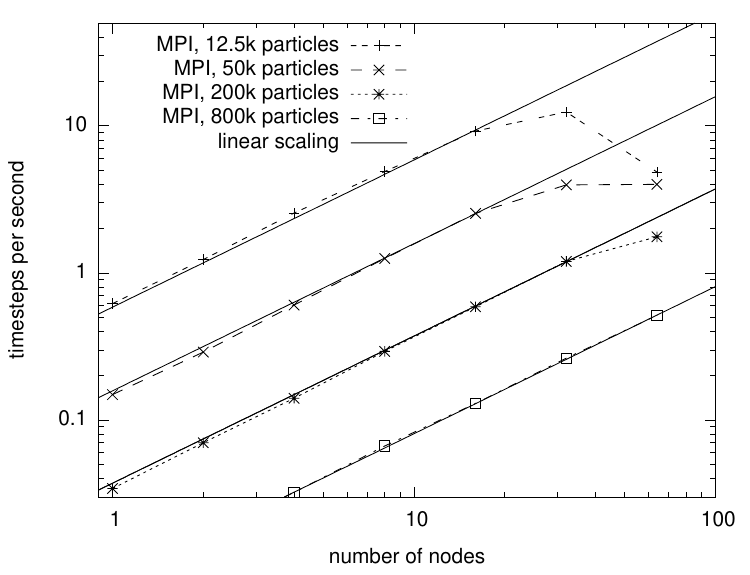}}
\label{fig:scalingstrong}
}
\subfigure[Weak scaling test: constant problem size per node.]{
\centering \resizebox{0.99\columnwidth}{!}{\includegraphics{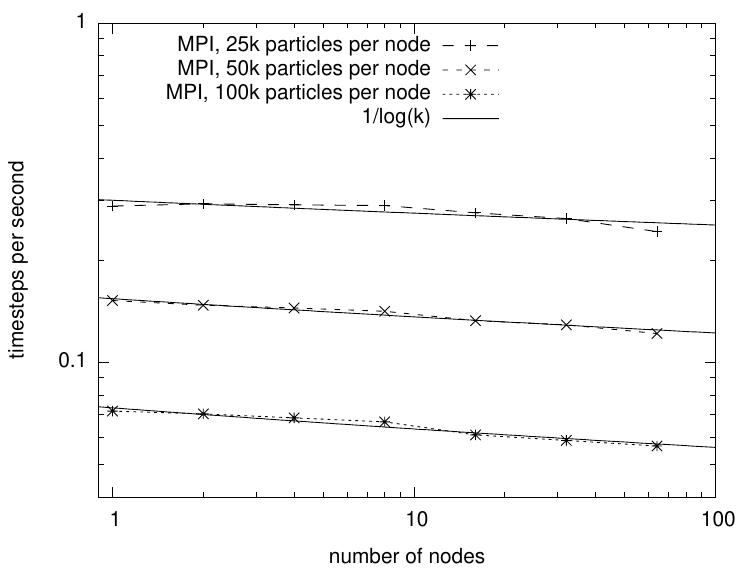}}
\label{fig:scalingweak}
}
\caption{Strong and weak scaling tests using a shearing sheet configuration with the \texttt{gravity\_tree.c} and \texttt{collisions\_tree.c} modules.}
\end{figure*}
Using the shearing sheet configuration with the tree modules \texttt{gravity\_tree.c} and \texttt{collisions\_tree.c}, we measure the scaling of \reb and the efficiency of the parallelization.
The simulation parameters have been chosen to resemble those of a typical simulation of Saturn's A-ring with an optical depth of order unity and a collision time-scale being much less than one orbit.
The opening angle is $\theta_{\mathrm{crit}} = 0.7$.
The \texttt{problem.c} files for this and all other tests can be found in the \texttt{test/} directory.

All scaling tests have been performed on the IAS aurora cluster. 
Each node has dual quad-core 64-bit AMD Opteron Barcelona processors and 16 GB RAM. 
The nodes are interconnected with 4x DDR Infiniband. 

\subsection{Strong scaling}\label{sec:scalingstrong}
In the strong scaling test the average time to compute one time-step is measured as a function of the number of processors for a fixed total problem size (e.g. fixed total number of particles).  
We use only the MPI parallelization option.

The results for simulations using $N=12.5\mathrm{k},50\mathrm{k},200\mathrm{k}$ and $800\mathrm{k}$ particles are plotted in Fig.~\ref{fig:scalingstrong}. 
One can see that for a small number of processors the scaling is linear for all problems.
When the number of particles per processor is below a critical value, $N_{pp}\sim2000$, the performance drops. 
Below the critical value, a large fraction of the tree constitutes the essential tree which needs to be copied to neighboring nodes every time-step.
This leads to an increase in communication.

The results show that we can completely utilize 64 processors cores with one million particles.

\subsection{Weak scaling}
In the weak scaling test we measure the average time to compute one time-step as a function of the number of processors for a fixed number of particles per processor.  
Again, we only use the MPI parallelization option.

The results for simulations using $N_{pp}=25\mathrm{k},50\mathrm{k}$ and $100\mathrm{k}$ particles per processor are plotted in Fig.~\ref{fig:scalingweak}. 
One can easily confirm that the runtime for a simulation using $k$ processors is $O(N_{pp} \log(N_{pp}\,k))$, as expected. 
By increasing the problem size, the communication per processor does not increase for the collision detection algorithm as only direct neighbors need to be evaluated on each node.
The runtime and communication for the gravity calculation is increasing logarithmically with the total number of particles (which is proportional to the number of processors in this case).

These results shows that \reb's scaling is as good as it can possibly get with a tree code.
The problem size is only limited by the total number of available processors.

\subsection{OpenMP/MPI trade-off}
\begin{figure}
\centering \resizebox{0.99\columnwidth}{!}{\includegraphics{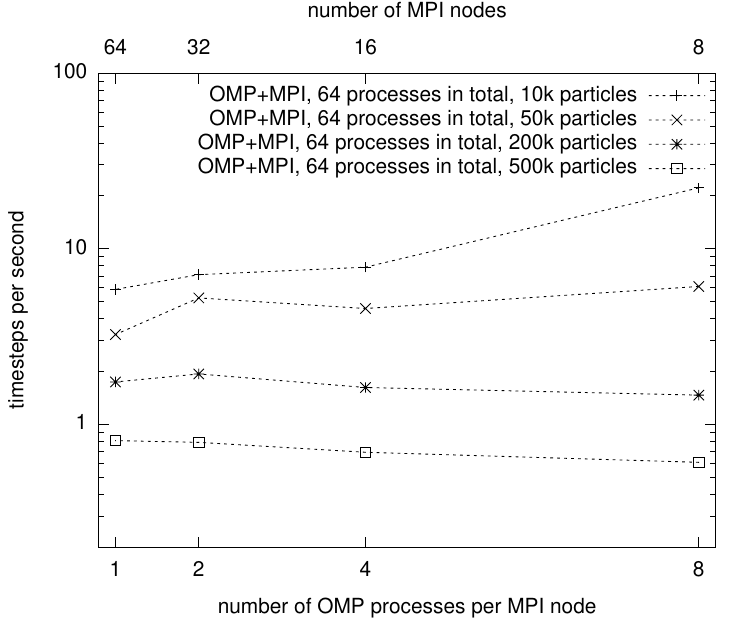}}
\caption{Comparison between OpenMP and MPI. Each run uses 64 CPU cores. A shearing sheet configuration the with \texttt{gravity\_tree.c} and \texttt{collisions\_tree.c} modules is used.\label{fig:scalingompmpi}}
\end{figure}
The previous results use only MPI for parallelization. 
\reb also supports parallelization with OpenMP for shared memory systems.

OpenMP has the advantage over MPI that no communication is needed. 
On one node, different processes share the same memory and work on the same tree and particle structures. 
However, the tree building and reconstruction routines are not parallelized.
These routines can only be parallelized efficiently when a domain decomposition is used (as used for MPI, see above).

Results of hybrid simulations using both OpenMP and MPI at the same time are shown in Fig.~\ref{fig:scalingompmpi}.
We plot the average time to compute one time-step as a function of the number of OpenMP processes per MPI node.
The total number of particles and processors~(64) is kept fixed. 

One can see that OpenMP does indeed perform better than MPI when the particle number per node is small and the run-time is dominated by communication (see also Sect.~\ref{sec:scalingstrong}). 
For large particle numbers, the difference between OpenMP and MPI is smaller, as the sequential tree reconstruction outweighs the gains.
Eventually, for very large simulations ($N_{pp}\gtrsim5000$) the parallelization with MPI is faster.

Thus, in practice OpenMP can be used to accelerate MPI runs which are bound by communication.
It is also an easy way to accelerate simulations on desktop computer which have multiple CPU cores. 

\subsection{Comparison of collision detection algorithms}\label{sec:scalingcollisions}
\begin{figure*}
\centering 
\subfigure[Varying the size of the simulation box and keeping a constant aspect ratio.]{
\resizebox{0.99\columnwidth}{!}{\includegraphics{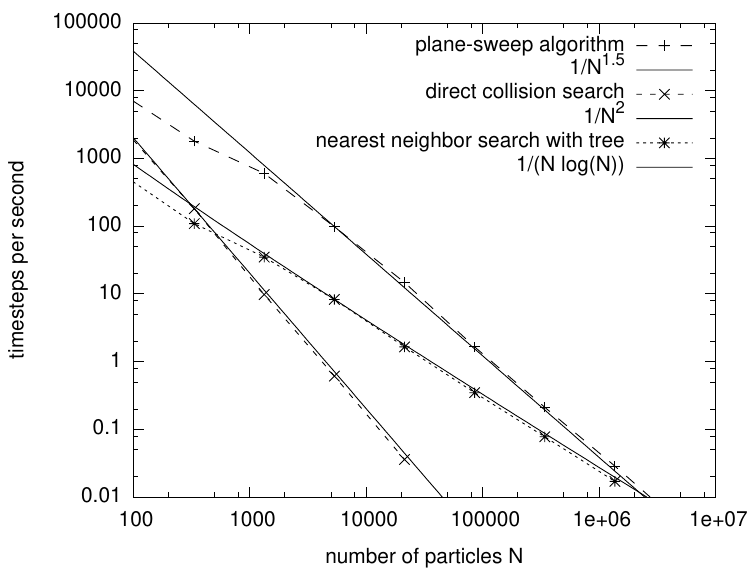}} \label{subfig:col}
}
\subfigure[Varying the radial size of the simulation box and keeping a constant azimuthal width.]{
\resizebox{0.99\columnwidth}{!}{\includegraphics{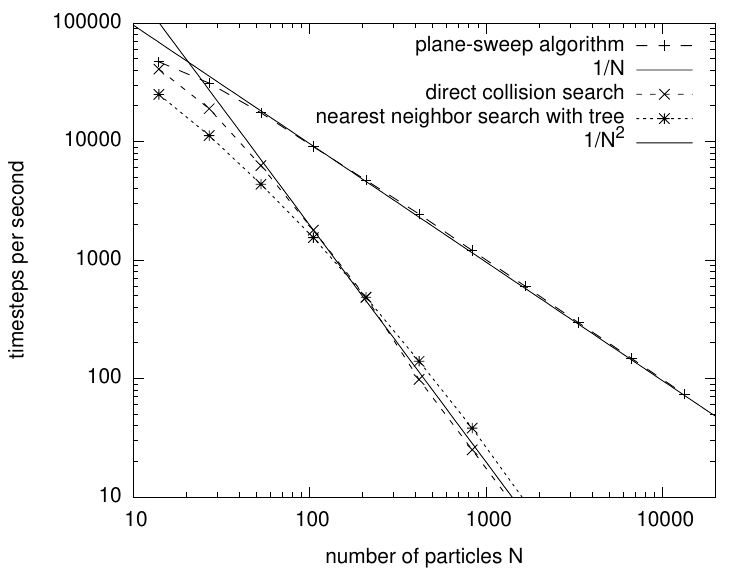}} \label{subfig:colelong}
}
\caption{Scalings of the plane-sweep algorithm, the octree and direct nearest neighbor search as a function of particle number. A shearing sheet configuration without self-gravity is used. \label{fig:scalingcollisions}}
\end{figure*}
The collision modules described in Sect.~\ref{sec:collisions} have very different scaling behaviors and are optimized for different situations. 
Here, we illustrate their scalings using two shearing sheet configurations with no self-gravity. 
We plot the average number of time-steps per second as a function of the problem size in Fig.~\ref{fig:scalingcollisions} for the plane-sweep algorithm and both the octree and direct nearest neighbor collision search. 

In simulations used in Fig.~\ref{subfig:col}, we vary both the azimuthal size, $L_y$, and radial size, $L_x$, of the computational domain. 
The aspect ratio of the simulation box is kept constant.
For the plane-sweep algorithm, the number of particle trajectories intersecting the plane\footnote{Note that a disk is effectively a two dimensional system. In three dimensions $N_\mathrm{SWEEPL} \sim L_y L_z\sim {N}^{2/3}$.} scales as $N_\mathrm{SWEEPL} \sim L_y\sim\sqrt{N}$. 
Thus, the overall scaling of the plane-sweep method is $O(N^{1.5})$, which can be verified in Fig.~\ref{subfig:col}.
Both the tree and direct detection methods scale unsurprisingly as $O(N\log(N))$ and $O(N^2)$, respectively.

For simulations used in Fig.~\ref{subfig:colelong}, we vary the radial size of the computational domain and keep the azimuthal size fixed at 20~particle radii. 
Thus, the aspect ratio changes and the box becomes very elongated for large particle numbers.
If a tree is used in \reb, an elongated box is implemented as many independent trees, each being a cubic root box (see Sect.~\ref{sec:domain}).
Because each tree needs to be accessed at least one during the collision search, this makes the tree code scale as $O(N^2)$ for large $N$, effectively becoming a direct nearest neighbor search. 
The plane-sweep algorithm on the other hand scales as $O(N)$, as the number of particle trajectories intersecting the plane is constant, $N_\mathrm{sweep} \sim L_y = const$.
Again, the direct nearest neighbor search scales unsurprisingly as $O(N^2)$.

From these test cases, it is obvious that the choice of collision detection algorithm strongly depends on the problem.
Also note that if the gravity module is using a tree, the collision search using the same tree comes at only a small additional cost.

The plane-sweep module can be faster for non-self-gravitating simulations by many orders of magnitude, especially if the problem size is varied only in one dimension.

\section{Summary}\label{sec:summary}
In this paper, we presented \reb, a new open-source multi-purpose N-body code for collisional dynamics. 
\reb is available for download at \url{http://github.com/hannorein/rebound} and can be redistributed freely under the GPLv3 license.

The code is written in a modular way, allowing users to choose between different numerical integrators, boundary conditions, self-gravity solvers and collision detection algorithms.
With minimal effort, one can also implement completely new modules.

The octree self-gravity and collision detection modules are fully parallelized with MPI and OpenMP. 
We showed that both run efficiently on multi-core desktop machines as well as on large clusters.
Results from a weak scaling test show that there is no practical limit on the maximum number of particles that \reb can handle efficiently except by the number of available CPUs. 
We will use this in future work to conduct extremely elongated simulations that can span the entire circumference of Saturn's rings. 

Two new collision detection methods based on a plane-sweep algorithm are implemented in \reb. 
We showed that the plane-sweep algorithm scales linearly with the number of particles for effectively low dimensional systems and is therefor superior to a nearest neighbor search with a tree.
Examples of effectively low dimensional systems include very elongated simulation domains and narrow rings. 
Furthermore, the simpler data-structure of the plane-sweep algorithm makes it also superior for quasi-two dimensional simulations with less than about one million particles.

Three different integrators have been implemented, for rotating and non-rotating frames.
All of these integrators are symplectic. 
Exact long-term orbit integrations can be performed with a Wisdom-Holman mapping.

Given the already implemented features as well as the open and modular nature of \reb, we expect that this code will find many applications both in the astrophysics community and beyond.
For example, molecular dynamics and granular flows are subject areas where the methods implemented in \reb can be readily applied.
We strongly encourage users to contribute new algorithms and modules to \reb.

\begin{acknowledgements}
We would like to thank the referee John Chambers for helpful comments and suggestions.
We would also like to thank Scott Tremaine, Hiroshi Daisaka and Douglas Lin for their feedback during various stages of this project.
Hanno Rein was supported by the Institute for Advanced Study and the NSF grant AST-0807444.
Shang-Fei Liu acknowledges the support of the NSFC grant 11073002.
Hanno Rein and Shang-Fei Liu would further like to thank the organizers of ISIMA~2011 and the Kavli Institute for Astronomy and Astrophysics in Beijing for their hospitality. 
\end{acknowledgements}

\bibliographystyle{aa}
\bibliography{full}

\end{document}